\documentclass[aps,prl,reprint,amsmath,amssymb,superscriptaddress,longbibliography,floatfix]{revtex4-2}
\usepackage[T1]{fontenc}
\usepackage{bm}
\usepackage{booktabs}
\usepackage{microtype}
\usepackage{etoolbox}
\usepackage{needspace}
\usepackage{pgfplots}
\pgfplotsset{compat=1.18}
\usepackage[hidelinks]{hyperref}
\hypersetup{pdfauthor={Arash Arabi Ardehali},
 pdftitle={Perturbative Holonomy Potentials and Vacua in Chiral Supersymmetric Gauge Theories}}

\newcommand{\cN}{\mathcal N}
\newcommand{\cF}{\mathcal F}
\newcommand{\bB}{\overline B}
\newcommand{\bK}{\overline K}
\newcommand{\eff}{\mathrm{eff}}
\newcommand{\out}{\mathrm{out}}
\newcommand{\inn}{\mathrm{in}}
\newcommand{\sgn}{\operatorname{sign}}
\newcommand{\RR}{\mathbb R}
\newcommand{\ZZ}{\mathbb Z}
\newcommand{\U}{\mathrm U}
\newcommand{\SU}{\mathrm{SU}}
\newcommand{\arxiv}[1]{\href{https://arxiv.org/abs/#1}{arXiv:#1}}

\newcounter{manuscriptbibliography}
\AtBeginEnvironment{thebibliography}{\stepcounter{manuscriptbibliography}}
\makeatletter
\AtBeginDocument{%
 \patchcmd{\NAT@bibsetnum}{LastBibItem}
   {LastBibItem\arabic{manuscriptbibliography}}{}{}%
 \patchcmd{\@bibsetup}{LastBibItem}
   {LastBibItem\arabic{manuscriptbibliography}}{}{}%
 \patchcmd{\endthebibliography}{LastBibItem}
   {LastBibItem\arabic{manuscriptbibliography}}{}{}%
}
\makeatother

\begin{document}

\title{Perturbative Holonomy Potentials and Vacua in Chiral Supersymmetric Gauge Theories}
\author{Arash Arabi Ardehali}
\affiliation{Department of Physics, Sharif University of Technology,
P.O. Box 11155-9161, Tehran, Iran}

\begin{abstract}
We present a Wilsonian framework, globally controlled across the holonomy space, for determining the perturbative data governing the vacua of chiral supersymmetric gauge theories on a circle. This allows lifting the Intriligator--Seiberg classification of semiclassical vacua in 3d $\mathcal{N}=2$ Chern--Simons-matter theories to $\RR^3\times S^1$. The framework is showcased by finding a rich set of vacuum sectors in a simple $\U(1)$ model and by reproducing on flat space the R-twisted reduction (first diagnosed on $S^3\times S^1$) of an $\SU(2)$ Maruyoshi--Song theory to the Gang--Yamazaki theory.
\end{abstract}
\maketitle

\emph{Introduction.---}Gauge holonomy around a circle changes the
Kaluza--Klein (KK) spectrum and can distinguish low-energy sectors.
With supersymmetric boundary conditions, the one-loop
Gross--Pisarski--Yaffe holonomy energy cancels between superpartners
\cite{GPY,Poppitz}. In chiral theories, however, perturbative
flatness does not follow: induced Fayet--Iliopoulos (FI) and
Chern--Simons (CS) couplings produce terms linear in the auxiliary
vector-multiplet $D$ fields, and a potential after these fields are
eliminated. Chirality of the KK spectrum may arise from the gauge
representations or from asymmetrically imposed flavor- or R-twisted boundary conditions around the circle.

Such D-terms are familiar in circle reductions, but a common
approach is indirect \cite{Aharony2013,Amariti2015}: one reduces
a nonchiral theory near trivial holonomy and then
asymmetrically decouples chirals by real-mass deformations. This
need not capture sectors elsewhere on the compact holonomy space.
A closed-form KK sum for the induced D-term coefficient near holonomy
saddles was obtained in the torus-reduction analysis of
Ref.~\cite[Sec.~3.3]{HLY}.
An independent derivation on
$\RR^3\times S^1$, including nonzero 4d FI
parameters, is given in the companion paper
\cite{Companion}.
This coefficient provides the starting point for a global
description of perturbative vacua throughout holonomy space.

This Letter develops that description jointly with
Ref.~\cite{Companion}. Combining the closed-form D-coefficient
with the Wilsonian decomposition of Ref.~\cite{AH}, we control
the leading holonomy potential in the outer patch of the holonomy space, where charged modes exceed
a matching cutoff, and determine the FI and CS couplings of
theories retaining light modes elsewhere. With a
sufficiently small cutoff, perturbative topological vacua lie in the
outer patch; Coulomb branches can extend into inner patches,
which may also contain semiclassical Higgs sectors. Infrared dynamics close to Coulomb-branch walls is outside the
focus of the Letter, but is supplied in the examples below to complete the otherwise partial picture provided by the perturbative matching. The companion
gives detailed derivations, higher-rank examples, and connections with
partition-function asymptotics \cite{Companion}. Nonperturbative monopole F-terms \cite{Aharony2013}, are beyond the
scope of this work; their analogous global treatment will appear in \cite{AARZ}.

\emph{Outer potential and its control.---}Consider an anomaly-free
four-dimensional $\cN=1$ gauge theory of rank $r_G$, weakly coupled
at $1/R_1$, where $2\pi R_1$ is the circle circumference. We consider flavor- or R-twisted supersymmetric
boundary conditions around the circle. On the
\emph{Coulomb slice}, charged scalar expectation values vanish;
neutral scalars are fixed to satisfy the tree-level F-term equations.
We assume no additional quadratic superpotential masses for charged
fields on this slice. Write the holonomy as $\exp(2\pi i\bm x)$,
with $\bm x$ in the Cartan torus modulo the Weyl group, and set
$\bm\sigma=\bm x/R_1$. The chiral KK masses are
\begin{equation}
 M_{n,\rho}=\frac{n+u_\rho}{R_1},\qquad
 u_\rho=\rho\cdot\bm x+a_\chi,\qquad n\in\ZZ,
 \label{eq:masses}
\end{equation}
where $\rho$ is a weight of the gauge representation of the multiplet $\chi$ and  $a_\chi$ its flavor or R-symmetry twist. Vector modes
associated with a root $\alpha$ have masses
$(n+\alpha\cdot\bm x)/R_1$.

A \emph{massless threshold} is where an additional charged KK
mode has zero real mass (at $D=0$); its codimension-one components
are \emph{walls}. Matter walls obey $u_\rho\in\ZZ$, and vector
walls obey $\alpha\cdot\bm x\in\ZZ$. Their complement consists
of \emph{chambers}. Choose a matching cutoff
$\Lambda_M=\epsilon/R_1$, with $0<\epsilon\ll1$ \cite{AH}.
The \emph{outer patch} has $|M|\geq\Lambda_M$ for every charged
mode. Its complement is covered by \emph{inner patches} with
fixed sets of retained light modes, $|M|<\Lambda_M$.

Let $D_i$ be the Cartan auxiliary fields and $F_i$ their linear
coefficient, normalized by
$\mathcal L_D=\tfrac12H_{ij}D_iD_j-F_iD_i/(4\pi)+O(D^3)$
\cite{IntriligatorSeiberg}, where $H_{ij}$ is the gauge kinetic
matrix. For renormalized 4d FI inputs
$\zeta_i^{(4)}$, set $\eta_i=(2\pi R_1)^2\zeta_i^{(4)}$;
the inherited 3d FI coupling is $\eta_i/R_1$
\cite{Companion}. Then
\begin{equation}
\begin{aligned}
 F_i^\out(\bm\sigma)
   &=\zeta_i^\eff+k_{ij}^\eff\sigma_j,\\
 \cF_i^\out(\bm x):={}&R_1F_i^\out\big(\frac{\bm x}{R_1}\big)=\eta_i-\frac12\sum_\rho\rho_i\bB_2(u_\rho),
\end{aligned}
\label{eq:main}
\end{equation}
where $\bB_2(u)=\{u\}^2-\{u\}+1/6$. A single charged mode
contributes $\delta F_i=\rho_i|M|/2$
\cite{DoreyTong,Tong2000,IntriligatorSeiberg} and the KK sum uses zeta-regularization:
$\sum_{n\in\ZZ}^{\zeta}|n+u|=-\bB_2(u)$ \cite{HLY,Companion}.
Paired vector modes cancel in $F_i$, although they enter the patch
definition and correct $H_{ij}$. The regulator and renormalized
FI input specify the matching scheme
\cite{DK,CorvilainGrimmRegalado,PW}.

Eliminating $D_i$ in the quadratic truncation gives
\begin{equation}
\begin{aligned}
 V_{D,\mathrm{quad}}
   =\frac{F_i^\out(H^{-1})_{ij}F_j^\out}{32\pi^2},\ \ \ 
 V_D^{(0)}=\frac{e_0^2\,|\bm\cF^\out|^2}{32\pi^2R_1^2}.
\end{aligned}
\label{eq:potential}
\end{equation}
The second expression uses $H_{ij}^{(0)}=e_0^{-2}\delta_{ij}$;
a general positive tree kinetic matrix replaces
$e_0^2\delta_{ij}$ by $(H^{(0)})^{-1}_{ij}$.
At $\bm\eta=0$, $F_i^\out$ is induced at one loop, so its square
gives a two-loop-order holonomy potential of order $g_4^2/R_1^3$.
Nonzero 4d FI parameters also give tree-level terms. 

The companion manuscript \cite{Companion} shows that the 4d weak-coupling hierarchy $g_4^2(1/R)\ll\epsilon^2\ll1$ controls the corrections to \eqref{eq:potential}
uniformly on the outer patch and keeps $H$ positive and nonsingular. Semiclassical vacua on the outer patch are therefore reliably captured by
\begin{equation}
    \boxed{\cF_i^\out(\bm x)=\eta_i-\frac12\sum_\rho\rho_i\bB_2(u_\rho)=0.}
\end{equation}

Cubic gauge-anomaly cancellation,
$\sum_\rho\rho_i\rho_j\rho_k=0$, makes $\bm\cF^\out$
affine on each chamber. Away from walls,
\begin{equation}
 k_{ij}^\out
 =\frac{\partial\cF_i^\out}{\partial x_j}
 =-\sum_\rho\rho_i\rho_j\bB_1(u_\rho),
 \label{eq:cs}
\end{equation}
where $\bB_1(u)=\{u\}-1/2$ off integers.
The function $\bm\cF^\out$ is continuous, periodic under
allowed large gauge transformations, and Weyl covariant.
Its slopes are piecewise constant CS levels that can jump
at walls. The FI and CS matching coefficients are one-loop
exact, unlike $H$ \cite{IntriligatorSeiberg}.

\emph{Inner couplings.---}Let $L$ label the chiral weights
with a retained charged mode on an inner patch. For
$\epsilon<1/2$, each weight has at most one such mode,
with KK number $-\operatorname{nint}(u_\rho)$, where
$\operatorname{nint}$ denotes the nearest integer. Set
$\delta(u)=u-\operatorname{nint}(u)$.
The retained mass is $M_\rho=\delta(u_\rho)/R_1$.
Subtracting only the retained mode from
each tower gives
\begin{equation}
\begin{aligned}
 \cF_i^\inn(\bm x)
   =\eta_i-\frac12\sum_{\rho\notin L}\rho_i\bB_2(u_\rho)-\frac12\sum_{\rho\in L}\rho_i\bK_2(u_\rho).
\end{aligned}
\label{eq:innF}
\end{equation}
The modified periodic Bernoulli functions
\cite[Sec.~2.2]{AH} are
\begin{equation}
\begin{split}
 \bK_2(u)&=\bB_2(u)+|\delta(u)|,\\
 \bK_1(u)&=\bB_1(u)+\tfrac12\sgn\delta(u),
\end{split}
\label{eq:Kfunctions}
\end{equation}
with $\sgn0=\bB_1(\mathbb{Z})=0$. For $|\delta|<1/2$, they reduce to
$\bK_2=\delta^2+1/6$ and $\bK_1=\delta$, smooth through
the threshold.

Choose a wall point $\bm x_*$ and local coordinates
$\bm s=(\bm x-\bm x_*)/R_1$. The local couplings are
\begin{equation}
\begin{aligned}
 k_{ij}^\inn
   &=-\sum_{\rho\notin L}\rho_i\rho_j\bB_1(u_\rho)
     -\sum_{\rho\in L}\rho_i\rho_j\bK_1(u_\rho),\\
 \zeta_{i,\mathrm{loc}}
   &=\cF_i^\inn(\bm x_*)/R_1,\\
 F_i^\inn(\bm s)
   &:=\frac{\cF_i^\inn(\bm x_*+R_1\bm s)}{R_1}
     =\zeta_{i,\mathrm{loc}}+k_{ij}^\inn s_j.
\end{aligned}
\label{eq:innercouplings}
\end{equation}
Anomaly cancellation again makes $k_{ij}^\inn$ constant
on the patch. At vector walls,
the light vectors must also be restored and the Cartan
coefficients assembled into couplings of the unbroken
gauge group, with its global form inherited from four
dimensions.

The matching across patches is explicit:
\begin{equation}
\begin{split}
 \cF_i^\out(\bm x)
   &=\cF_i^\inn(\bm x)
     +\frac12\sum_{\rho\in L}\rho_i|\delta(u_\rho)|,\\
 k_{ij}^\out
   &=k_{ij}^\inn
     +\frac12\sum_{\rho\in L}\rho_i\rho_j\sgn M_\rho.
\end{split}
\label{eq:thresholdmatching}
\end{equation}
Integrating out
retained modes where perturbatively controlled recovers
the outer coefficients; changing $\epsilon$ only
redistributes contributions between explicit fields and
loop sums.
Two simple consequences of \eqref{eq:thresholdmatching} are: $i$) if all retained chirals are massless at $\bm x_*$, then
$\bm\cF^\inn(\bm x_*)=\bm\cF^\out(\bm x_*)$;
$ii$) for a single wall, away from intersections,
$k^\inn$ is the average of the adjacent outer levels.

For an abelian inner theory with canonically normalized
light scalars, the classical vacuum equations are (cf.~\cite{IntriligatorSeiberg})
\begin{equation}
\boxed{\begin{aligned}
 \cF_i^\inn(\bm x_*+R_1\bm s)&=R_1F_i^\inn(\bm s)=2\pi R_1\sum_{a\in L}\rho_{a,i}|\phi_a|^2,\\
 M_a(\bm s)\phi_a&=0,\qquad
 \partial_{\phi_A}W_{\mathrm{loc}}=0.
\end{aligned}}
\label{eq:wallvacua}
\end{equation}
Here $M_a(\bm s)=M_a(\bm0)+\rho_a\cdot\bm s$, and $A$
runs over all retained chirals, including neutral ones.
Solutions are taken mod gauge transformations.
The matched superpotential includes tree-level elimination
of heavy chirals but no loop corrections
\cite{Rocek,SeibergNR,Weinberg}. For a nonabelian inner theory, the
matter moment map replaces the scalar sum
\cite{IntriligatorSeiberg}.

A condensate lies on its own matter wall,
$M_a(\bm s)=0$. If all condensing fields are massless
at $\bm s=0$, their charges constrain $\bm s$ to their
common kernel. In rank one this pins $s=0$; in higher
rank, unlifted directions can support mixed
Higgs--Coulomb branches.

At fixed charges, multiplicities, and Chern--Simons levels,
a sufficient semiclassical hierarchy for suppressed gauge loops in an abelian Higgs
sector is
\begin{equation}
\begin{gathered}
 e_0^2\ll m_{H,\ell},\quad
 (m_H^2)_{ij}\sim e_0^2
 \sum_{a:\phi_a\neq0}
 \rho_{a,i}\rho_{a,j}|\phi_a|^2,
\end{gathered}
\label{eq:wallhierarchy}
\end{equation}
where $m_{H,\ell}^2$ are the positive eigenvalues of the
parity-even Higgs mass-squared matrix.
For nonabelian inner theories, the bound applies to the
square roots of the positive eigenvalues of the full
parity-even gauge mass-squared matrix \cite{Companion}. Local control also requires a regular,
perturbatively controlled K\"ahler description with weak
superpotential interactions. Any residual topological gauge
sector must decouple from the light chirals or have
independently controlled IR dynamics.

If they do not condense,
retained charged chirals can have $M_a(\bm s)\neq0.$ On the Coulomb slice of an abelian inner patch, if every retained charged chiral satisfies
$e_0^2\ll|M_a|$, integrating out then corrects $F^\inn$ as:
\begin{equation}
\begin{aligned}
 F_i^\inn(\bm s)
 +\frac12\sum_{a\in L}\rho_{a,i}|M_a(\bm s)|=\frac{\cF_i^\out(\bm x_*+R_1\bm s)}{R_1}.
\end{aligned}
\label{eq:innerCoulomb}
\end{equation}
Thus D-flatness requires $\bm\cF^\out=0$, even when
$\bm F^\inn\neq0$: retained-field loops can cancel
the matched D-coefficient at $\phi_a=0$.
This recovers the continued outer D-coefficient by
further integration within the inner theory. If only
some fields are integrated out, only their contributions
are added to $\bm F^\inn$.

\emph{Classification of perturbative vacua.---}
At fixed $a_\chi$ and $\eta_i$, choose $\epsilon$
small enough that every isolated off-wall zero of
$\bm\cF^\out$ lies in the outer patch, and then take
the weak-coupling limit. All perturbative topological
vacua considered here are therefore outer. An isolated
off-wall zero has nonsingular $k^\out$ and gives a
gapped abelian Maxwell--CS gauge sector. If
$k^\out$ has a kernel, any zero extends locally along
it; these directions and their dual photons parametrize
a perturbative Coulomb branch
\cite{IntriligatorSeiberg}.

Inner patches contain controlled extensions of Coulomb
branches through Eq.~\eqref{eq:innerCoulomb}, and
semiclassical Higgs or mixed Higgs--Coulomb sectors
through Eq.~\eqref{eq:wallvacua}, with retained-field
corrections included as needed. 

Sufficiently close to
walls, the relevant masses can approach $e_0^2$ and
the gauge dynamics can become strong. This infrared (IR)
regime, including residual interacting massless sectors,
is outside our perturbative analysis. Due to strong coupling, a wall zero alone
does not establish a quantum vacuum; see the second example below.

Lifting by nonperturbative monopole F-terms is completely beyond the scope of this work. The examples below do not have such corrections though.

Two examples with vacuum sectors at nontrivial holonomy demonstrate the stakes of global matching next: the first with a nonzero threshold value of the Coulomb-slice D-coefficient canceled by a charged condensate for $\eta\neq0$; the second with strong-coupling IR lifting of the trivial-holonomy sector despite its vanishing D-coefficient.

\begin{table}[htbp]
\caption{Positive isolated roots $x_i^0$ at $\eta=0$ and
their Cartan CS levels $k_i$. For small
nonzero $\eta$, they move to $x_i^0-\eta/k_i$ without
changing levels. Negative roots have opposite levels.}
\label{tab:abelian}
\centering
\setlength{\tabcolsep}{4pt}
\renewcommand{\arraystretch}{1.3}
\begin{tabular}{c*{6}{c}}
\toprule
$x_i^0$ & $\frac3{26}$ & $\frac6{35}$ & $\frac{17}{71}$ &
 $\frac{13}{42}$ & $\frac{14}{39}$ & $\frac{15}{32}$\\
$k_i$ & $78$ & $-35$ & $71$ & $-42$ & $39$ & $32$\\
\bottomrule
\end{tabular}
\end{table}

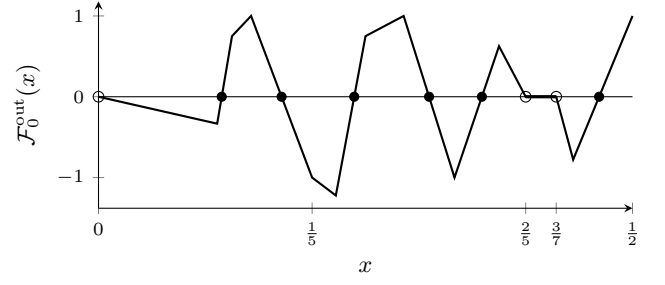
\begin{figure}[htbp]
\centering
\begin{tikzpicture}
\begin{axis}[
 width=\columnwidth,height=1.7in,
 xlabel={$x$},ylabel={$\cF^\out_0(x)$},
 xmin=0,xmax=0.5,ymin=-1.38,ymax=1.18,
 xtick={0,0.2,0.4,0.428571428571,0.5},
 xticklabels={$0$,$\frac15$,$\frac25$,$\frac37$,$\frac12$},
 ytick={-1,0,1},
 tick label style={font=\scriptsize},
 label style={font=\small},
 axis lines=left,cycle list={{solid}},clip=false]
\addplot[thin,no marks] coordinates {(0,0) (0.5,0)};
\addplot[thick,no marks] coordinates {
 (0,0) (0.111111111111,-0.333333333333)
 (0.125,0.75) (0.142857142857,1)
 (0.2,-1) (0.222222222222,-1.222222222222)
 (0.25,0.75) (0.285714285714,1)
 (0.333333333333,-1) (0.375,0.625)
 (0.4,0) (0.428571428571,0)
 (0.444444444444,-0.777777777778) (0.5,1)};
\addplot[only marks,mark=*,mark size=1.7pt] coordinates {
 (0.115384615385,0) (0.171428571429,0)
 (0.239436619718,0) (0.309523809524,0)
 (0.358974358974,0) (0.46875,0)};
\addplot[very thick,no marks] coordinates {(0.4,0) (0.428571428571,0)};
\addplot[only marks,mark=o,mark size=2pt] coordinates {
 (0,0) (0.4,0) (0.428571428571,0)};
\end{axis}
\end{tikzpicture}
\caption{The outer coefficient $\cF^\out$ (continued to the inner patches through \eqref{eq:abelian}) of the abelian model
at $\eta=0$; reflection gives the negative-$x$ half.
Filled points are isolated outer vacua, the bold segment
is a Coulomb interval, and open circles are wall zeros.
For nonzero $\eta$, Coulomb-slice zeros intersect the line
$\cF^\out_0=-\eta$. GPT-6 Astra assisted with the plotting code.}
\label{fig:abelian}
\end{figure}

\needspace{3\baselineskip}
\emph{An abelian model and its FI deformation.---}Consider
a $\U(1)$ theory with periodic chirals of charges
\begin{equation}
 (q_1,q_2,q_3,q_4,q_5)=(1,5,-7,-8,9),\qquad W=0,
 \label{eq:charges}
\end{equation}
satisfying $\sum_iq_i=\sum_iq_i^3=0$. The theory is
understood below its ultraviolet cutoff, with weak
coupling at $1/R_1$. For $-1/2<x\leq1/2$,
\begin{equation}
 \cF^\out_\eta(x)=\eta-\frac12\sum_iq_i\bB_2(q_ix),
 \label{eq:abelian}
\end{equation}
where $\eta=(2\pi R_1)^2\zeta^{(4)}$. Note that $\cF^\out_0$
is even and its chamber slopes are odd. At $\eta=0$,
\begin{equation}
 \{x:\cF^\out_0(x)=0\}
 =\{0\}\cup\{\pm x_i^0\}_{i=1}^{6}\cup\pm[\tfrac25,\tfrac37],
 \label{eq:abelianzeros}
\end{equation}
with the isolated roots and levels given in
Table~\ref{tab:abelian}; see Fig.~\ref{fig:abelian}.
The twelve isolated roots support gapped Maxwell--CS
sectors. On the two open intervals,
$k^\out=\zeta^\out=0$, so the holonomy and dual photon
parametrize one-complex-dimensional perturbative
Coulomb branches.

At $x=0$, all five zero-mode chirals remain light,
with $k^\inn=0$ and $F^\inn=\eta/R_1$. The classical
Higgs locus $2\pi\sum_iq_i|\phi_i|^2=\eta/R_1$ modulo
$\U(1)$ has generic complex dimension four.
The Supplemental Material presents a candidate
interacting superconformal field theory (SCFT) at its strongly-coupled
origin for $\eta=0$.

The endpoints $x_L=2/5$ and $x_R=3/7$ retain
\begin{equation}
\begin{array}{c|cc}
 &x_L&x_R\\ \hline
\text{light theory}&\U(1)_{-25/2}+\Phi_{+5}
                  &\U(1)_{-49/2}+\Phi_{-7}\\
\zeta_{\mathrm{loc}}&\eta/R_1&\eta/R_1
\end{array}.
\label{eq:u1walls}
\end{equation}
Their negative-holonomy partners have opposite levels
but the same light charges and FI parameters.
At $\eta=0$, known infrared dualities identify the
endpoint theories with $\ZZ_5$ and $\ZZ_7$ gaugings
of a free chiral \cite[Sec.~4.1]{IntriligatorSeiberg},
with the finite gaugings fixed by the inherited
unit-charge normalization. These are
infrared results beyond our perturbative framework, added here for completeness.

The Coulomb branch nevertheless extends perturbatively
into the inner patch. At $\eta=0$, with $s=(x-x_L)/R_1$,
\begin{equation}
\begin{aligned}
 F^\inn(s)&=-\frac{25}{2}s,\qquad M_{+5}=5s,\\
 F^\inn(s)+\frac52|M_{+5}|
   &=-\frac{25}{2}s+\frac{25}{2}|s|.
\end{aligned}
\label{eq:u1innermatching}
\end{equation}
For $s>0$ and $25e_0^2\ll M_{+5}$,
the retained chiral can be integrated out perturbatively.
Its loop cancels $F^\inn$ and its CS shift cancels
$k^\inn$, giving a Coulomb vacuum with $\phi_{+5}=0$
despite $F^\inn\neq0$. The matching at $x_R$ is similar.

For fixed $0<|\eta|<1/3$, the isolated roots shift
and new ones appear, giving sixteen topological sectors
for $\eta>0$ and fourteen for $\eta<0$.
With the cutoff chosen as above, these roots are outer.
There are also endpoint Higgs vacua at
$x=\pm x_L$ for $\eta>0$ and $x=\pm x_R$ for
$\eta<0$, leaving $\ZZ_5$ or $\ZZ_7$ unbroken.
At sufficiently weak coupling their masses satisfy
Eq.~\eqref{eq:wallhierarchy}. No other wall with a
single light chiral supports a semiclassical Higgs
vacuum in this FI range. The endpoints thus realize
a second cancellation mechanism: their nonzero $F^\inn$ is balanced by scalar condensates.

\begin{table}[t]
\caption{Matched data at every wall of the twisted
$(A_1,A_2)$ example. Subscripts denote Cartan charges;
$W_\pm$ have charges $\pm2$. The endpoints restore
$\SU(2)$ and the interior walls retain $\U(1)$.
Levels are supersymmetric Wilsonian coefficients.}
\label{tab:ad}
\centering
\setlength{\tabcolsep}{3.5pt}
\renewcommand{\arraystretch}{1.3}
\begin{tabular}{lccccc}
\toprule
$x_*$ & $0$ & $\frac1{10}$ & $\frac15$ & $\frac25$ & $\frac12$\\
\midrule
Light fields & $W_\pm$ & $\phi_{-2}$ & $q'_{+1}$ &
 $q_{-1},\phi_{+2}$ & $W_\pm$\\
$R_1\zeta_{\mathrm{loc}}$ & $0$ & $\frac15$ & $0$ & $-\frac15$ & $0$\\
$k_{\U(1)}^\inn$ & $2$ & $0$ & $-\frac32$ & $\frac12$ & $2$\\
$k_{\SU(2)}^\inn$ & $1$ & --- & --- & --- & $1$\\
\bottomrule
\end{tabular}
\end{table}

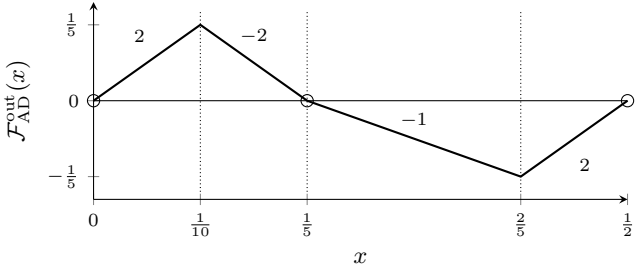
\begin{figure}[t]
\centering
\begin{tikzpicture}
\begin{axis}[
 width=\columnwidth,height=1.65in,
 xlabel={$x$},ylabel={$\cF_{\mathrm{AD}}^\out(x)$},
 xmin=0,xmax=0.5,ymin=-0.26,ymax=0.26,
 xtick={0,0.1,0.2,0.4,0.5},
 xticklabels={$0$,$\frac1{10}$,$\frac15$,$\frac25$,$\frac12$},
 ytick={-0.2,0,0.2},
 yticklabels={$-\frac15$,$0$,$\frac15$},
 tick label style={font=\scriptsize},label style={font=\small},
 axis lines=left,cycle list={{solid}},clip=false]
\addplot[thin,densely dotted,no marks]
 coordinates {(0.1,-0.24) (0.1,0.24)};
\addplot[thin,densely dotted,no marks]
 coordinates {(0.2,-0.24) (0.2,0.24)};
\addplot[thin,densely dotted,no marks]
 coordinates {(0.4,-0.24) (0.4,0.24)};
\addplot[thin,no marks] coordinates {(0,0) (0.5,0)};
\addplot[thick,no marks] coordinates {
 (0,0) (0.1,0.2) (0.2,0) (0.4,-0.2) (0.5,0)};
\addplot[only marks,mark=o,mark size=2.3pt] coordinates {
 (0,0) (0.2,0) (0.5,0)};
\node[font=\scriptsize] at (axis cs:0.043,0.17) {$2$};
\node[font=\scriptsize] at (axis cs:0.15,0.17) {$-2$};
\node[font=\scriptsize] at (axis cs:0.30,-0.05) {$-1$};
\node[font=\scriptsize] at (axis cs:0.46,-0.17) {$2$};
\end{axis}
\end{tikzpicture}
\caption{Continued outer coefficient of the nonabelian
example. Numbers are outer Cartan levels. Open circles
mark wall zeros, not established vacua; the nonzero
wall values also enter through the local FI couplings
in Table~\ref{tab:ad}. GPT-6 Astra assisted with the plotting code.}
\label{fig:ad}
\end{figure}

\emph{An R-twisted nonabelian example.---}Consider the
Maruyoshi--Song Lagrangian flowing to the $(A_1,A_2)$
theory \cite{MaruyoshiSong}, treating the free factors
as in Ref.~\cite{Bridging}. Its charged matter is an
adjoint $\phi$ and two doublets $q,q'$ of $\SU(2)$.
The minimal $\U(1)_r$ twist \cite{Dedushenko,Bridging} is
\begin{equation}
\begin{gathered}
 \varphi(y+2\pi R_1)=e^{i\pi(\mathsf F+2r)}\varphi(y),\\[2pt]
 \begin{array}{c|ccc}
 \chi&q&q'&\phi\\ \hline
 \rho&\pm1&\pm1&0,\pm2\\
 a_\chi&2/5&-1/5&1/5
 \end{array},
\end{gathered}
\label{eq:adinput}
\end{equation}
where $\mathsf F$ is fermion number and $r$ is the
$\cN=2$ $\U(1)_r$ charge. This is a second-sheet
reduction in index terminology \cite{CassaniKomargodski}.
Neutral chirals have nonintegral shifts and do not
remain light. Matching in the weakly coupled
Lagrangian gives Fig.~\ref{fig:ad} on the Weyl alcove
$0\leq x\leq1/2$ and the wall data in
Table~\ref{tab:ad}. There are no outer zeros.

At $x_*=1/5$, only $q'_{+1}$ becomes massless.
For $s=(x-1/5)/R_1$, the continued outer coefficient is
\begin{equation}
 F^\out(s):=\frac{\cF_{\mathrm{AD}}^\out(1/5+R_1s)}{R_1}
 =-\frac32s+\frac12|s|.
 \label{eq:admatching}
\end{equation}
Subtracting the light chiral contribution gives
$F^\inn=-3s/2$, hence
\begin{equation}
 \U(1)_{-3/2}+\Phi_{+1},\qquad
 \zeta_{\mathrm{loc}}=0,\qquad W_{\mathrm{loc}}=0.
 \label{eq:gy}
\end{equation}
This is the
Gang--Yamazaki gauge--matter sector \cite{GangYamazaki},
previously obtained from the index \cite{Bridging}.
Background contact terms require separate matching
\cite{Bridging}, and infrared supersymmetry enhancement
\cite{GangYamazaki} is additional strong-coupling information beyond our perturbative framework.

At $x_*=0,1/2$, the restored $\SU(2)$ theories have
$k_{\U(1)}^\inn=2$. In the fundamental charge
normalization $\pm1$, $k_{\U(1)}=2k_{\SU(2)}$
\cite{IntriligatorSeiberg}, so both are
pure $\cN=2$ $\SU(2)_1$. Their known strong-coupling infrared dynamics
has no supersymmetric vacuum
\cite{IntriligatorSeiberg}.
These wall zeros therefore illustrate the limitation
of the perturbative criterion \eqref{eq:wallvacua}.

The remaining walls have nonzero FI terms. At
$x=1/10$, the light chiral has charge $-2$, so its
moment map cannot balance
$\zeta_{\mathrm{loc}}=1/(5R_1)>0$. At $x=2/5$,
the inherited interaction $W\supset\phi qq$
\cite{MaruyoshiSong} gives
$W_{\mathrm{loc}}\propto\phi_{+2}q_{-1}^{\,2}$.
Its F-term forces $q_{-1}=0$, leaving a positive-charge
moment map unable to balance
$\zeta_{\mathrm{loc}}=-1/(5R_1)$.
Neither wall supports a semiclassical Higgs vacuum.
The global calculation thus recovers the known
interacting sector while distinguishing semiclassical
exclusions from conclusions requiring strong-coupling infrared dynamics.

\emph{Outlook.---}The one-loop-exact D-coefficient and
Wilsonian threshold subtraction give a global
perturbative description of holonomy potentials and
vacua as the light spectrum changes. Outer vacua
follow without a kinetic-matrix calculation; inner
FI and CS couplings determine where loops or
condensates can cancel the matched D-term.
Strong-coupling infrared dynamics near walls and
nonperturbative F-terms require separate analysis.

The framework makes it possible to revisit
earlier works on circle-compactified chiral theories, such as Ref.~\cite{LeeTerning}, in which D-term
effects were not incorporated throughout the holonomy
space, and to analyze
on flat space further R-twisted reductions of
Maruyoshi--Song-type theories \cite{MS,AMS} whose
supersymmetry-enhancing reductions have so far been only
diagnosed on $S^3\times S^1$ \cite{GKS,Bridging,AGRS}.

\begin{acknowledgments}
\textit{Acknowledgments.}---The author thanks A.~Amariti,
C.~Closset, E.~Poppitz, and A.~Zanetti for related discussions,
and especially D.~Resnick for collaboration on the higher-rank
applications in the companion paper \cite{Companion}.
OpenAI GPT-6 Astra assisted with manuscript organization and
editing, literature searches, calculation checks, exploration of all walls in the two examples, and LaTeX
and plotting code. The
ancillary Mathematica notebook was
generated through prompts guiding an emphasis on the subtle role of the
emergent 3d symmetry (the $a$-direction), then
verified, polished, and revised block by block.
\end{acknowledgments}

\newpage

\section*{Supplemental Material: Infrared phase of the abelian wall theory}

\emph{Setup.---}We study the wall theory at $x=0$ of the Letter's
abelian example at zero FI parameter. All five zero-mode chirals
remain light, giving $\U(1)_0$ with charges $Q=(1,5,-7,-8,9)$ and no FI parameter.
We take $W=0$, with no additional UV-induced polynomial or
monopole superpotential. The compact $\U(1)$ and flux lattice
$m\in\ZZ$ retain the parent theory's unit-charge normalization.
This supplemental infrared analysis begins from, but goes beyond, the
perturbative matching in the Letter. We find a self-consistent
$F$-maximum and a nontrivial index. Their interpretation as
evidence for an interacting SCFT assumes the stated interactions
and the absence of omitted accidental R-mixing symmetries.

\emph{Flavor symmetry and R-mixing.---}The origin theory has
five matter phase rotations modulo the gauged combination,
and the topological current $j_J=(2\pi)^{-1}\star da$.
Its continuous non-R flavor algebra is therefore
$\mathfrak u(1)^4\oplus\mathfrak u(1)_J$
\cite{1IntriligatorSeiberg}. For scalar R-charges $r_i$,
we fix the gauge-R mixing scheme
$r_i\mapsto r_i+\lambda Q_i$ by setting $Q\cdot r=0$
and use the four-dimensional anomaly-free reference
\begin{equation}
\begin{gathered}
 r^{(0)}=\left(\frac32,\frac{14}{15},\frac{43}{42},1,1\right),\\
 \sum_iQ_i^2(r_i^{(0)}-1)=0.
\end{gathered}
 \label{eq:reference}
\end{equation}
Three inherited nonanomalous flavor directions are
\begin{equation}
\begin{aligned}
 f_1&=(-1,-2,2,-2,1),\\
 f_2&=(-7,6,3,-2,-2),\\
 f_3&=(-8,-2,-7,5,1).
\end{aligned}
\label{eq:flavors}
\end{equation}
They satisfy $Q\cdot f_A=\sum_iQ_i^2f_{A,i}=0$.
The full three-dimensional trial assignment is
\begin{equation}
 r=r^{(0)}+\sum_{A=1}^3t_Af_A+a(1,1,1,1,1),
 \label{eq:trial}
\end{equation}
together with topological mixing $\tau$. The additional
matter direction obeys $Q\cdot\boldsymbol1=0$ but has
four-dimensional anomaly $220a$. Thus the reference
descends from four dimensions, whereas the infrared
mixing space need not. Removing this symmetry would
require an interaction absent from our setup. In
particular, the usual dynamically generated monopole
superpotentials that eliminate anomalous four-dimensional
symmetries after reduction \cite{1Aharony2013} are absent
for this abelian parent theory.

\emph{Sphere partition function.---}In radius-one units,
\begin{equation}
 Z_{S^3}(r,\tau)=\int_{\RR}d\sigma\,
 e^{2\pi\tau\sigma+\sum_i\ell(1-r_i+iQ_i\sigma)},
 \label{eq:sphere}
\end{equation}
with $F=-\log|Z_{S^3}|$, $\ell'(z)=-\pi z\cot(\pi z)$,
and $\ell(0)=0$. The superconformal R-symmetry locally
maximizes $F$ over conserved flavor directions
\cite{1Jafferis,1ClossetFmax}. Writing
$S(r)=\sum_i|Q_i|(1-r_i)$, the real contour converges
absolutely for $S(r)>2|\tau|$. Outside this domain,
the restricted calculation uses the pole-avoiding contour
deformation described in the ancillary notebook.

Restricting Eq.~\eqref{eq:trial} to $a=0$ yields a
stationary point with a positive Hessian eigenvalue,
hence a saddle rather than an $F$-maximum. We therefore
use the full $W=0$ mixing space, including $a$.
Extremizing over $(t_1,t_2,t_3,a,\tau)$ gives
\begin{equation}
\begin{aligned}
 r_*\simeq{}&(0.49761224,\,0.45297353,\,0.41829870,\\
            &\phantom{(}0.39721876,\,0.37148456),\\
 \tau_*={}&0,\qquad F_*\simeq4.282376845.
\end{aligned}
\label{eq:rstar}
\end{equation}
All five physical Hessian eigenvalues are negative.
The axial mixing is 
\begin{equation}
    a_*\simeq-0.6007689223\neq0,
\end{equation}
so $r_*$ is not a four-dimensional
anomaly-free assignment.

The lowest-R polynomial gauge invariants and their
dimensions are
\begin{equation}
\begin{aligned}
 \mathcal O_A&=\Phi_{+5}\Phi_{-7}^{\,2}\Phi_{+9},
 &\Delta_A&\simeq1.66105549,\\
 \mathcal O_B&=\Phi_{+1}^{\,2}\Phi_{+5}\Phi_{-7},
 &\Delta_B&\simeq1.86649670.
\end{aligned}
\label{eq:operators}
\end{equation}
Neither reaches the free-chiral bound. The bare monopole
vacuum has gauge charge $3|m|$ and R-charge
$8.89415903\ldots|m|$, and requires electric dressing
\cite{1IntriligatorSeiberg}. These checks support the
extremization under the stated assumptions about accidental
symmetries: no gauge-invariant scalar operator has R-charges below $1/2$ to violate unitarity.

\emph{Superconformal index.---}With flavor fugacities
set to one, the index is the protected trace over
states on $S^2$,
\begin{equation}
 \mathcal I(q)=
 \operatorname{Tr}_{\Delta-R-j_3=0}
 (-1)^{\mathsf F}q^{(\Delta+j_3)/2},\qquad |q|<1.
 \label{eq:trace}
\end{equation}
A scalar chiral primary contributes $q^{R/2}$.
In the holomorphic-determinant convention, abelian
localization gives \cite{1ImamuraYokoyama,1Bridging}
\begin{equation}
\begin{split}
 \mathcal I(q)={}&\sum_{m\in\ZZ}\oint\frac{dz}{2\pi iz}\,
 e^{i\pi k^+_{gg}m}z^{k^+_{gg}m}q^{k^+_{gR}m/2}\\
 &\times\prod_i
 \frac{(z^{-Q_i}q^{-Q_im/2+1-r_i/2};q)_\infty}
      {(z^{Q_i}q^{-Q_im/2+r_i/2};q)_\infty},
\end{split}
\label{eq:index}
\end{equation}
where $(u;q)_\infty=\prod_{n\geq0}(1-uq^n)$ and
\begin{equation}
\begin{aligned}
 k^+_{gg}&=k_{gg}+\frac12\sum_iQ_i^2,\\
 k^+_{gR}&=k_{gR}+\frac12\sum_iQ_i(r_i-1).
\end{aligned}
 \label{eq:levels}
\end{equation}
The unshifted levels are supersymmetric Wilsonian
coefficients; background fluxes vanish and the
vacuum contributes one.

For the wall theory at $x=0$, $k_{gg}=k_{gR}=0$,
giving $k^+_{gg}=110$ and $k^+_{gR}=0$ in our gauge-R mixing scheme. Reorganizing the determinants in terms
of $|Q_im|$ gives the vacuum factor
$(-1)^m z^{3|m|}q^{A_*|m|}$, where
$A_*=S(r_*)/4=4.44707952\ldots$.
All remaining letter powers are positive. Hence the
low-order index comes entirely from $m=0$; taking
the constant term in $z$ gives
\begin{equation}
\begin{split}
 \mathcal I_*(q)={}&1+q^{\Delta_A/2}+q^{\Delta_B/2}-5q\\
 &+q^{1.068793895\ldots}+q^{1.087219023\ldots}\\
 &+O(q^{11/10}).
\end{split}
\label{eq:originseries}
\end{equation}
The exponents are numerical evaluations of exact
linear combinations of $r_i^*$. The coefficient $-5$
is consistent with four ordinary flavor currents
and the topological current, with no marginal
chiral monomial at this assignment \cite{1FazziEtAl}.
Together with Eq.~\eqref{eq:rstar}, the nontrivial
index supports an interacting SCFT under the stated
assumptions.

The ancillary notebook \texttt{u1\_ir\_supplement.nb}
gives the stationarity and Hessian calculations,
contour and resolution checks, the bare monopole gauge and R charges, operator enumeration,
and the index expansion.

\end{document}